# AGGREGATED EVALUATION OF OPERATION QUALITY OF COMPLEX HIERARCHICAL NETWORK SYSTEMS

**Olexandr Polishchuk**

Department of Nonlinear Mathematical Analysis, Pidstryhach Institute for Applied Problems of Mechanics and Mathematics, National Academy of Sciences of Ukraine, Lviv, Ukraine, od_polishchuk@ukr.net

**Abstract.** The main approaches for the formation of generalized conclusions about operation quality of complex hierarchical network systems are analyzed. Advantages and drawbacks of the "weakest" element method and a weighted linear aggregation method are determined. Nonlinear aggregation method is proposed for evaluating the quality of the system, which consists of elements of the same priority. Hybrid approaches to form generalized conclusions are developed based on the main aggregation methods. It is shown that they allow to obtain more reliable aggregation results.

**Key words:** complex system, network, hierarchy, priority, evaluation, aggregation

## 1. Introduction

Complex dynamical systems (CDS) are used almost in all areas of human activity, e.g. in transportation (railway, road and aviation systems, transportation networks of large cities and regions of countries) [1], supply and logistics (systems for power, gas, petrol, heat and water supply, trade networks) [2], information and communication (Internet, TV, radio, post service, press, fixed and mobile telephony) [3], in economics (networks of state-owned and (or) private companies, their suppliers and final products distributors) [4], finance (banking and insurance networks, money transfer systems) [5], education, healthcare etc. Their state and operation quality impose large impact on citizens' quality of life, efficiency of economy and possibilities for its development, as well as government structures readiness to mitigate impacts of technological and natural disasters [6]. Finally, they may be treated as the evidences of country development level in general [7]. Failure of one of the elements of such systems can often lead to operation breakdown or destabilization of the whole CDS. The example of this is cascading phenomenon [8]. Often the situations of the kind (e.g. accidents at nuclear or large chemical plants and other hazardous facilities, power lines, gas pipelines etc.) may lead to harsh consequences, such as environmental disasters, property loss and numerous human victims [9]. These circumstances determine the importance of continuous monitoring of technological systems operation, careful control of their behavior and timely response to emerging threats. Solution for this problem belongs to the fields of systems theory, system analysis, complex networks theory, mathematical modelling etc [10-12]. It is usually difficult to implement classical mathematical modelling methods on practice for studying most existing CDS due to the problems of dimension and adequacy. Therefore, decomposition and investigation of the properties of the separate components of the system is the usual method for modeling. However, the important features of the interaction between components of CDS may be lost during decomposition. A lot of problems arise when trying to optimize the complex large scale systems [13].

Optimizing separate components does not always guarantee the quality improvement of the whole system, and synchronous optimization of all components of CDS is usually unrealistic goal (hard to imagine the simultaneous repair of all roads of megapolis). Network analysis methods [11, 14] are focused mostly on studying of system structure and interconnections between network elements without analysis of their state and functioning quality. At the same time, flow processing in the node may be quite a complicated process [15]. System analysis in general is aimed on selection of alternatives about further actions with respect to real CDS (development plans, ways to optimize, etc.) [16]. Taking into account the human factor is a separate problem in systems research. Often the influence of this factor can not account for any mathematical methods.

Complex systems appear, operate and develop within long periods of time and with natural processes of "aging", despite regular improvements, more strict and accurate control over their behaviour is required. This is why the development of methods for evaluation and forecasting the state, operation quality and interaction between structural elements of CDS is actual problem [17-21]. Often the cause of accidents is the wrong evaluation of the current situation or an inadequate forecast of its future development. Sometimes the reason of accidents become "unfavorable coincidence of many random unlikely circumstances". Scientific discipline is not created yet, which could mathematically formalize such causes and forestall the disasters (accidents on the railway, nuclear power plants, hazardous industries, etc.). Catastrophe theory [22], which includes the bifurcation theory of dynamical systems (differential equations) and the theory of singularities of smooth mappings, is still far from solving the problems of real systems. Evaluation theory allows to determine the preconditions that can lead to catastrophic events (deterioration of the state or operation quality of system components, the weakening of the interaction between them, etc.). It also allows to determine the elements that threaten or require urgent optimization, and to analyze the influence of this optimization on the other system components and the whole CDS.

Multi-criteria and multi-parameter analysis of the state and operation quality of the system elements leads to a huge number of local evaluations. Manual analysis of such amounts of data to make correct and timely decisions is practically impossible. Therefore, a special place in evaluation theory take methods for formation of generalized (aggregated) conclusions.

Aggregation of evaluations of system components is known for a long time [21, 23, 24]. We used it to make generalized conclusions about the behavior of system elements according to the set of characteristics, parameters and evaluation criteria [25, 26] in a continuous, discrete, conceptual, and precise rating scale [20, 21]. Aggregation is useful for determining the optimal operation modes of CDS [27] and selecting an optimal system from a given class of equivalent systems [28]. It can be used to analyze the history of CDS functioning and forecasting its behavior [20, 29].

As a rule, the weighted linear aggregation method (WLAM) is used to form generalized conclusions about the state and operation quality of the system [23-26]. In this paper we analyze three approaches to form aggregated evaluations: the "weakest" element method (WEM), WLAM and nonlinear aggregation method (NAM). We determine the relationship between the evaluations obtained by these methods and propose hybrid approaches for aggregated evaluation of systems that increase adequacy of generalized conclusions.

## 2. Evaluation of Complex Hierarchical Network Systems

The most of created and controlled by a man industrial, transportation, financial and other systems have a hierarchical network structure. Complex hierarchical network systems (CHNS) are special because each subsystem of a certain hierarchy level consists of a set of subsystems which form subnetwork of lower hierarchy level network (see Fig. 1). Every hierarchical level of such system is the collection of nodes, connected by edges through which the flows are passing. The edges shall ensure smooth passage of flow and nodes are to ensure its processing. Hierarchy is introduced on the basis of management system construction principles, CDS objects arrangement in space etc. Flow movement for which CHNS was created is performed at network of the lowest level. At the higher (control) levels, flows are represented by information, organizational and administrative decisions etc.

Methods for complex evaluation of the state, operation quality and interaction between components of CHNS were described in [31]. These methods detemine the way for reflecting CHNS experimental studies data onto structured, according to hierarchy, sequence of local, forecasting, interactive and generalized conclusions about system behaviour. Taking into account the diversity of CHNS objects, these methods defines universal principles of such evaluations development, common for all objects of the same type and functional destination considering peculiarities of the former.

Real network systems can be combined into more complex structures: multiplexes, associations, and conglomerates [14, 30]. Analysis of the interaction of these formations is also an important issue.

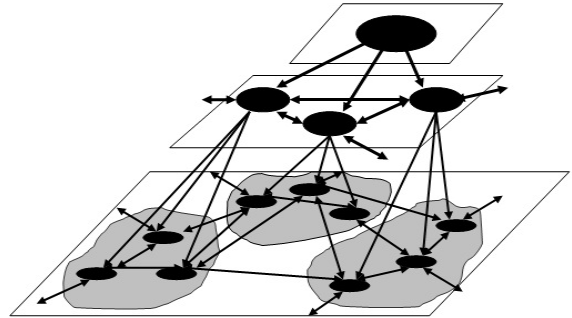

Fig. 1 Hierarchical network structure

Evaluation of real systems is reasonable to start with objects of lowest structural level, i.e. with their elements. We define an element as an object of clearly defined location, functional destination and relevant set of characteristics describing its state and functioning process with corresponding ranges of permissible values for those characteristics. All characteristics are evaluated according to certain collection of criteria and parameters [21]. Of course, evaluation of every object presupposes evaluation of its state on the first place, and only after that the evaluation of quality of implementation of its functions that in any case depend on element's state – either directly or indirectly. The process of evaluation is started only after the stage of thorough extraction and processing of experimental data as to each of characteristic and their conversation into format, suitable for further analysis. Evaluations for elements' state and functions they implement on the basis of their characteristics behaviour analysis we call *local* [29].

Scheduled inspections of system's components are held at different time points, which means the results of last study may not stay on such stage till following inspection, and state of component and its operation quality may cross "safety threshold" [32]. It should be also taken into account that every real system evolves in time, i.e. with regard to current requirements, its evaluation may be insufficient. Therefore, evaluation process should contain means of analysis of CDS's meeting expected requirements for short- and long-term perspective. Thus, the evaluation process should not only determine conclusions and discover "faulty" elements for the time point moment when study is held, but also it should forecast further behaviour of system components. ***Forecasting*** analysis performed on the basis of local evaluations prehistory, allows us to determine the nature, direction and speed of system state change, follow up negative processes and forecast potential risks, as well as material and financial expenses required for their elimination or timely prevention [29].

Due to the number of reasons, scheduled inspections may often not discover drawbacks that arise "out of schedule". It should be also taken into account that even excellent state and functioning quality of separate components in the system do not ensure high performance of its subsystems or system in general. And vice versa, the most optimal work organization process will not ensure high efficiency of system functioning if CHNS's state or organization of components functioning is unsatisfactory. The more worn-out CHNS's objects are the more urgent is the problem of continuous monitoring of their state and functioning process. Quality of implementation of functions by component may be affected by number of third-party factors, both internal and external as to the system. Internal influence may be evaluated on the level of subsystems connecting interacting objects. We call this evaluation method *interactive* [33]. It allows us to determine separate components in selected subsystem, functioning of which is unsatisfactory, without thorough analysis of state and functioning quality of these components and expenses related to such analysis. The simplest interactive evaluation may be performed for system where the movement of flows is deterministic, at least partially, in accordance with certain schedule, the compliance to which may be periodically summed up [1].

In general, only if combined, proposed methods may provide sufficiently full and adequate understanding of CHNS quality. Indeed, high local evaluations do not ensure effective interaction of elements, failures of separate systems components may result in breakdown in balanced organization, satisfactory state of object for the moment of current inspection does not imply the state will stay satisfactory till the next inspection. Huge amount of information regarding separate CHNS elements without appropriate generalization is ill-suited for rapid analysis and timely reaction for drawbacks discovered. On higher generalization levels, evaluation allows to determine reliable conclusion as to the state and operation quality of system and its main subsystems and to define measures, as well as material and finance expenses required for its modernisation and optimization of functioning [34]. At the local level evaluation allows to identify separate elements and their components subject to improvement. These "narrow" places that are constantly discovered during scheduled inspections or continuous system monitoring may be subject to mathematical modelling. This narrows down the object of modelling and makes the process itself more realistic.

### 3. Priority of System Components

The concept of priority of system components and functions that they implement, is important in the study of CDS [4]. Priority of the component can be defined as a quantitative measure of its significance in the system (in every system there are more and less important objects). In order to reduce the dimension during simulation into the system content and structure first of all are included its highest priority components. In other cases after decomposition of CDS primary consideration is also given to the study of high-priority objects. The highest priority usually have components, failure of which leads to a failure of the whole system.

Priority of network node can determine its degree and/or betweenness centrality. Among two nodes with the same degree or betweenness centrality the highest priority has the node through which the more flows have movement. The last statement is true for the edges of the network system. Usually, the analysis of the effectiveness of the large scale network system operation can not be carried out simultaneously for all elements (nodes, edges, flows) and components of the higher hierarchy levels. Sometimes the efficiency of certain subsystems of CDS is more important than the operation quality of separate system elements (in human collectives is expressed by the term "team work": an agreed team of "mediocrities" often wins uncoordinated team of "stars"). Routes of flows may be such subsystems in CHNS. The priority of each route is determined by the number and/or volume of flows that pass this route in a certain time period. For flow which moves the route is not important priority of node in which it is delayed. This means that from the point of view of analysis of flows movement on the system all the nodes and edges of the route may be considered equally important. Routes can be grouped according to their priority.

### 4. Aggregation Methods: the Main Approaches

The main objective of aggregated evaluation is to create an adequate generalized puttern of the operation quality of system components of various hierarchy levels. Analysis of this puttern can be done by avoiding processing of huge volumes of information that describes the behavior of these components in detail. Aggregated evaluation should significantly simplify and shorten the process of problems localization and preparing of appropriate decisions. Next we analyze the adequacy of generalized conclusions obtained using three different approaches. We illustrate these approaches on the simplest example of a system $S$ that consists of $N$ elements $s_n$, $n = \overline{1,N}$. Let us assume that $e(s_n)$ is evaluation of operation quality of element $s_n$, $e(s_n) \in [e^{\min}, e^{\max}]$, where $e^{\min}$ and $e^{\max}$ are the minimum and maximum possible evaluations of the operation quality of element $s_n$ correspondingly. Methods of formation of multi-criteria and multi-parameter current and forecasting evaluations of behavior of system elements in precise ball scale were described in detail in [29].

Under the first approach the system quality $e^{(1)}(S)$ is determined by the quality of its "weakest" element, i.e.

$$e^{(1)}(S) = \min_{n=1,N} e(s_n).$$

This approach has to use when we evaluate the systems in which the malfunction of elements can lead to failure of the separate subsystems or the whole CDS. An example of such system is the human body in which the unsatisfactory functioning of certain organs can lead to lethal consequences; railway in which the shifts of subgrade may cause a train crash; the gas pipelines in which the result of a crack in the pipe often were explosions with a large number of human victims, etc. Unsatisfactory state of such elements may cause cascading effects [8] that extend in complex networks (massive power outage in regions of the country, spread of epidemics, occurrence of traffic jams, etc.). WEM is one of the main methods for determining the reliability of the technical and information systems. It is an ideal means for evaluation of conveyor systems work. If elements with the lowest evaluations may pose a threat to the functioning of CDS, the elements with the highest evaluations may be a "sample" and used to improve other elements of real systems [20]. Such elements can be used to determine the criteria of practically achievable optimality [26].

Another and the most common approach for obtaining generalized conclusions is a weighted linear aggregation of elements evaluations. In the case of the system considered above, this evaluation is obtained by the ratio

$$e^{(2)}(S, \boldsymbol{\rho}) = <\boldsymbol{\rho}, \mathbf{e}(\mathbf{s})>_{R^N} / <\boldsymbol{\rho}, \mathbf{1}>_{R^N},$$

where $\mathbf{e}(\mathbf{s}) = \{e(s_n)\}_{n=1}^N$, $\boldsymbol{\rho} = \{\rho_n\}_{n=1}^N$ is a vector of weighted coefficients, which determines the priority of system elements, $\mathbf{1} = \{1\}_{n=1}^N$, and $<.,.>_{R^N}$ is a scalar product in Euclid space $R^N$. The main disadvantage of WLAM is neglecting both positive and negative evaluations. Consider the following example. Assume for the production of some device requires supply of $N$ equally important components. Even in the case of delivery 100% of ($N-1$) components and 10% of $N$-th component can be product only 10% devices. At the same time, evaluation obtained by WLAM when $\boldsymbol{\rho} \equiv \mathbf{1}$, for the scale $[e^{\min}, e^{\max}]=[0, 100]$ gives the value $e^{(2)}(S,\mathbf{1}) = 100\text{-}90/N$. For large values of $N$, this evaluation is close to 100%, although it is possible to product only 10% of devices. It is obvious that such evaluation does not correspond to reality. This method is absolutely not fit for evaluating the operation of conveyor systems.

A more appropriate for the reality results of generalization are obtained using the method of non-linear aggregation. It is based on the following statement [35].

**Theorem 1.** For an arbitrary set of real numbers $\{a_n\}_{n=1}^N$, $a_n > 0$, such that $\sum_{n=1}^N a_n = A$, the maximum value $\prod_{n=1}^N a_n$ is achieved in the case of $a_n = a^* = A/N$, $n = \overline{1, N}$.

If $\{a_n\}_{n=1}^N$ is a set of evaluations, this means that the best result among all sets is achieved, when deviation of $a_n$, $n = \overline{1, N}$, from the mean value $a^*$ is a minimum. In the case of system $S$ considered above, evaluation of its quality by NAM is obtained by using the ratio

$$e^{(3)}(S) = \prod_{n=1}^N e(s_n) / (e^*)^{N-1},$$

where $e^* = \sum_{n=1}^N e(s_n)/N$. The main drawback, which significantly limits the use of the third approach, is the difficulty of taking into account the priority of system elements.

Generalization by means of NAM can be called the rule of "optimality of mediocrity". This rule works wonderfully on a conveyor belt, where from employees require the same productivity. Otherwise conveyor performance is equal to productivity of the "weakest" employee. For normal production requires timely supply of all components, to make the right decision must be synchronized delivery of information flows, etc. However, use this rule, for example, for evaluation of research team work is inexpedient. In this case usually "one person makes a discovery, the second person this discovery confirms (or refutes) and third person "washes the tubes". "Mediocrity" of all is counterproductive here. These considerations determine the areas of potential use of NAM.

Let us assume that $e(s_n) = e^{(1)}(S)$, $n = \overline{1, N}$. Then

$$e^{(3)}(S) = \frac{\prod_{n=1}^N e^{(1)}(S)}{(\sum_{n=1}^N e^{(1)}(S)/N)^{N-1}} = \frac{(e^{(1)}(S))^N N^{N-1}}{(\sum_{n=1}^N e^{(1)}(S))^{N-1}} =$$

$$= \frac{(e^{(1)}(S))^N N^{N-1}}{(e^{(1)}(S))^{N-1} N^{N-1}} = e^{(1)}(S).$$

Thus, for an arbitrary set of evaluations $\{e(s_n)\}_{n=1}^N$ such that $e(s_n) \geq e^{(1)}(S)$, $n = \overline{1, N}$, we obtain

$$e^{(3)}(S) \geq e^{(1)}(S).$$

Rewrite the aggregated nonlinear evaluation $e^{(3)}(S)$ in the form

$$e^{(3)}(S) = \frac{\prod_{n=1}^N e(s_n)}{(e^{(2)}(S,\mathbf{1}))^{N-1}}$$

and show that

$$e^{(3)}(S) \leq e^{(2)}(S,\mathbf{1})$$

or

$$\frac{\prod_{n=1}^N e(s_n)}{(e^{(2)}(S,\mathbf{1}))^{N-1}} \leq e^{(2)}(S,\mathbf{1}).$$

From the last inequality we obtain

$$\prod_{n=1}^N e(s_n) \leq (e^{(2)}(S,\mathbf{1}))^N = \frac{(\sum_{n=1}^N e(s_n))^N}{N^N}$$

or

$$N^N \prod_{n=1}^N e(s_n) \leq (\sum_{n=1}^N e(s_n))^N.$$

The validity of the last inequality follows from the well-known inequality [36]

$$\sqrt[N]{\prod_{n=1}^{N} a_n} \leq \frac{\sum_{n=1}^{N} a_n}{N}, \quad a_n > 0, \quad n = \overline{1,N}.$$

Furthermore, in the case $e(s_n) = e^*$ we obtain

$$N^N \prod_{n=1}^{N} e(s_n) = N^N \prod_{n=1}^{N} e^* = N^N (e^*)^N$$

and

$$(\sum_{n=1}^{N} e(s_n))^N = (Ne^*)^N = N^N (e^*)^N.$$

Thus, the next result is in order.

**Theorem 2.** For arbitrary set of evaluations $\{e(s_n)\}_{n=1}^{N}$, $e(s_n) \geq 0$, $n = \overline{1,N}$, and equally important elements of the system $S$ are fair inequalities

$$e^{(1)}(S) \leq e^{(3)}(S) \leq e^{(2)}(S, \mathbf{1}),$$

and $e^{(1)}(S) = e^{(3)}(S)$ if $e(s_n) = e^{(1)}(S)$, and $e^{(3)}(S) = e^{(2)}(S, \mathbf{1})$ if $e(s_n) = e^*$, $n = \overline{1,N}$.

Consider as an example the system $S = \{s_n\}_{n=1}^{N}$, $e(s_n) \in [0, 100]$, $n = \overline{1,N}$, $N = 3,4,5$. Take a limiting case when the evaluation $e(s_1)$ increases uniformly from 0 to 100 and the values $e(s_n) = 100$, $n = \overline{2(1)N}$. We suppose that all system elements have the same priority, i.e. $\boldsymbol{\rho} \equiv \mathbf{1}$. Graphs of behavior of aggregated evaluations obtained using the three approaches described above are shown in Fig. 2. We can see that NAM gives a more adequate and closer to the reality aggregated evaluations for the system of equally important elements than WLAM.

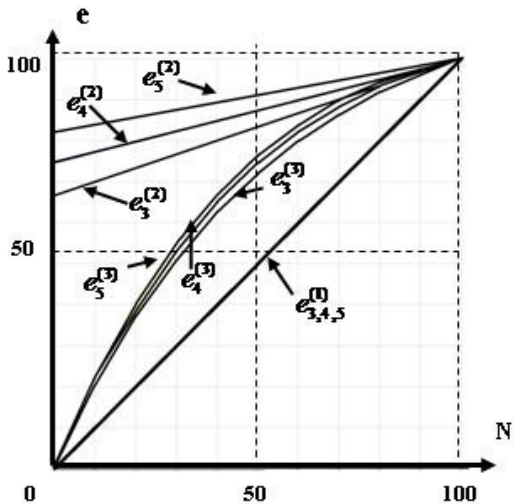

Fig. 2. Generalized conclusions obtained by means of WEM ($e^{(1)}_{3,4,5}$), WLAM ($e^{(2)}_{3,4,5}$) and NAM ($e^{(3)}_{3,4,5}$)

If operation quality of all system elements increases uniformly and simultaneously, the evaluations $e^{(2)}_{3,4,5}$ and $e^{(3)}_{3,4,5}$ are behaving as the evaluation $e^{(1)}_{3,4,5}$. In other words, the evaluations obtained by means of all three approaches converge to the same result, if all values $e(s_n)$, $n = \overline{1,N}$, are approaching to $e^*$.

Shown in the Fig. 2 results confirm that the detection and optimization of the "weakest" elements improve quality evaluation of the whole CDS. The methods discussed above should not be confused with the well-known weighted sum and weighted product methods [37]. These methods are expedient to use for determination of the best or the worst element (alternative) from a given set. In fact, these methods form a relative quality evaluation of one element in comparison with the other system elements. We aim to be based on absolute quality evaluations [29].

## 5. Hybrid Aggregation

It is advisable to combine the approaches described above to get more adequate generalized conclusions. Two ways of such combining can be determined. The first way is to remove the defects detected when using the first approach, and further use the second or third approach. In this case WEM is advisable to apply only to a group of elements with the highest priority.

The relative difference between the evaluations obtained by a combination of first and second or first and third approaches is a quantitative measure of the adequacy of the obtained generalized conclusion. For the above example of the production of some device when $N=3$ and $e(s_1) = 10\%$, $e(s_2) = e(s_3) = 100\%$, the evaluations obtained by WEM, WLAM and NAM in the percentage scale will have the following values: $e^{(1)}(S) \approx 10\%$, $e^{(2)}(S, \mathbf{1}) \approx 70\%$, $e^{(3)}(S) \approx 20.4\%$. When first and second approaches are combined the measure of adequacy of aggregated evaluations is determined by the ratio

$$\sigma_{1,2}(S) = (e^{(2)}(S, \mathbf{1}) - e^{(1)}(S))/e^{(2)}(S, \mathbf{1}).$$

The maximum value of this parameter (equal to 1) is achieved with the greatest deviation of aggregated linear evaluation from the evaluation of the "weakest" element and the minimum (equal to 0) value is achieved when the evaluations of all system elements are equal. For the last example we have $\sigma_{1,2}(S) \approx 0.857$.

When first and third approaches are combined the measure of adequacy of aggregated evaluations is determined by the ratio

$$\sigma_{1,3}(S) = (e^{(3)}(S) - e^{(1)}(S))/e^{(3)}(S).$$

The maximum value of this parameter (equal to 1) is achieved with the greatest deviation of aggregated nonlinear evaluation from the evaluation of the "weakest" element and the minimum (equal to 0) value is achieved when the evaluations of all system elements are equal. For the last example we have $\sigma_{1,3}(S) \approx 0.509$. Thus, NAM generates more adequate generalized conclusion than WLAM.

The second way for combination of aggregation methods is as follows. Assume that the elements of system $S$ can be divided into groups $S_m = \{s_n\}_{n=1}^{n_m}$, $n_m \geq 1$, $m = \overline{1,M}$, $\sum_{m=1}^{M} n_m = N$. Suppose that all elements of the group $S_m$ have the same priority $\widetilde{\rho}_m$, $m = \overline{1,M}$. Then the aggregated evaluation of system $S$ can be done using the ratio

$$e^{(4)}(S, \widetilde{\boldsymbol{\rho}}) = <\widetilde{\boldsymbol{\rho}}, \mathbf{e}^{(3)}(\mathbf{s})>_{R^M} / <\widetilde{\boldsymbol{\rho}}, \mathbf{1}>_{R^M},$$

where $\mathbf{e}^{(3)}(\mathbf{s}) = \{e^{(3)}(S_m)\}_{m=1}^{M}$ and $\widetilde{\boldsymbol{\rho}} = \{\rho_m\}_{m=1}^{M}$ is a vector of weighted coefficients, which determines the priority of groups $S_m$, $m = \overline{1,M}$. The last ratio determines a hybrid method for formation of aggregated evaluations that combines WLAM and NAM. It allows to take into account the priority of elements of separate groups (for example, traffic routes of flows) and perform a more adequate evaluation within each group.

From theorem 2 we have that the following statement is in order.

**Theorem 3.** The next inequalities are fair

$$e^{(1)}(S) \leq e^{(4)}(S, \widetilde{\boldsymbol{\rho}}) \leq e^{(2)}(S, \boldsymbol{\rho}^*),$$

where $\boldsymbol{\rho}^* = \{\{\rho_m\}_{n=1}^{n_m}\}_{m=1}^{M}$.

If all elements of the system $S$ have a different priority ($M=N$), the proposed hybrid method is converted into WLAM. If all elements of the system have the same priority ($M=1$), then it is converted into NAM.

Consider the following example. Assume that the system consists of two groups of elements. Three elements form a first group. Priority of elements of this group is equal to 1.0. The operation quality of the first element increases from 0 to 100, and the operation quality of other two elements of this group is equal to 100. Two elements form a second group. Priority of elements of this group is equal to 0.5. The operation quality of these elements is equal to 50. Using NAM in this case is not possible, since the elements of the system have different priority. Graphs of behavior of aggregated evaluations, obtained in this case by means of WEM, WLAM and by hybrid method that combines WEM and WLAM are shown in Fig. 3.

Note that the operation quality of the elements of the second group is not improved and continues to be equal to 50. Therefore, the aggregated evaluation of the system obtained by WEM is equal to 50 even if the operation quality of elements of the first group is equal to 100. This is another reason to use WEM only for a set of high-priority elements.

**6. Conclusions**

In general, complex large scale dynamical systems with various structures require different approaches to aggregation. These approaches should ensure maximum adequacy of generalized conclusions [38]. WLAM does not always guarantee this adequacy, because it can "hide" potentially dangerous components of the system. WEM can be too "radical" and give a negative conclusion about the system quality, based on the quality evaluations of the least important elements. NAM is useful if correlation of system component evaluations should be minimal. Thus, all considered above methods have limited areas of application. Using these methods without taking into account the specifics of the system and its separate components can lead to unreliable results of generalization. We propose to use the hybrid approaches for aggregation which should as much as possible to take into account this specificity. The foregoing examples confirm the effectiveness of such approaches.

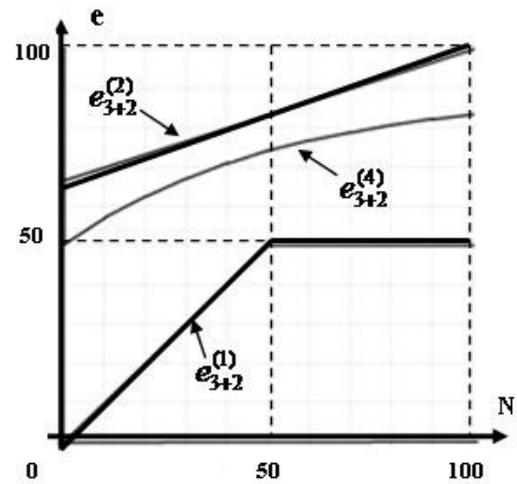

Fig. 3. Generalized conclusions obtained by means of WEM ($e_{3+2}^{(1)}$), WLAM ($e_{3+2}^{(2)}$) and hybrid aggregation method ($e_{3+2}^{(4)}$)